\documentstyle[aps,prc,preprint,tighten,epsf,dina4]{revtex}

\newcounter{teile}
\def\eqnar{\stepcounter{equation}
\setcounter{teile}{\value{equation}} \setcounter{equation}{0}
\renewcommand\theequation{\arabic{teile}\alph{equation}}
\begin{eqnarray}}
\def\endeqnar{\end{eqnarray}
\renewcommand\theequation{\arabic{equation}}
\setcounter{equation}{\value{teile}}\hspace*{-2ex}}
\newcommand{\braket}[2]{\left\langle{#1}\right.\left|{#2}\right\rangle}
\newcommand{\bramket}[3]{\left\langle\,{#1}\,\left|\,{#2}\,
            \right|\,{#3}\,\right\rangle}
\newcommand{\intp}[1]{\int\frac{d^{3}{#1}}{(2\pi)^{3}}}
\newcommand{\intpp}[1]{\int\frac{d^{4}{#1}}{(2\pi)^{4}}}
\newcommand{\bqn}{\begin{eqnarray}}
\newcommand{\eqn}{\end{eqnarray}}
\newcommand{\nn}{\nonumber\\}
\newcommand{\ov}{\overline}
\newcommand{\eps}{\epsilon}

\newcommand{\dreimat}[9]{ \left( \begin{array}{*{3}{c}}
                                      #1 & #2 & #3 \\
                                      #4 & #5 & #6 \\
                                      #7 & #8 & #9  
                               \end{array} \right) }
\newcommand{\zweivek}[2]{ \left( \begin{array}{c} #1 \\ #2 
                               \end{array} \right) }
\newcommand{\half}{\frac{1}{2}}
\newcommand{\ra}{\rightarrow}

\begin{document}

\title{A covariant diquark-quark model of the nucleon in the 
Salpeter approach}
\author{V. Keiner}
\address{Institut f\"ur Theoretische Kernphysik,\\
         Universit\"at Bonn, Nussallee 14-16, D-53115 Bonn, FRG}
\date{March 5, 1996}
\preprint{\vbox{Bonn TK-96-06}}
\maketitle

\begin{abstract}
We develop a rather simple, formally covariant quark-diquark model of 
the nucleon. The nucleon is treated as a bound state of a constituent
quark and a diquark interacting via a quark exchange. We include both
scalar and axial-vector diquarks. The underlying Bethe-Salpeter 
equation is transformed into a pair of coupled Salpeter equations. 
The electromagnetic form factors of the nucleon are calculated 
in the Mandelstam formalism. We obtain a very good description 
of all electromagnetic form factors for momentum transfers 
up to $- 3 \; (\mbox{GeV/c})^2$.
\end{abstract}

\vspace{1cm}

\vspace{4cm} 
\noindent
e-mail: {\em keiner@pythia.itkp.uni-bonn.de} \\
Tel.: +49 (0)228 73 2377 \\
Fax: +49 (0)228 73 3728 \\ 

\newpage

\section{Introduction}
In a recent paper we presented a simple, formally covariant 
scalar-diquark--quark model of the nucleon \cite{ke}. Combining the
Salpeter approach with the Mandelstam formalism we calculated the 
electromagnetic form factors of the nucleon. Despite the simplicity
of the model, we obtained a very good description of the measured
proton electric form factor for momentum transfers up to $- 3 \;
\mbox{GeV}^2$. However, the 
electric form factor of the neutron came out a factor of more than three
too large, and the calculated magnetic form factors failed to describe
the data quantitatively. Since scalar diquarks do not contribute to the
magnetic current density, the magnetic form factors result only from
the coupling to the quark. This is the main motivation for introducing also
spin 1 diquarks. Of course, in the non-relativistic quark model, 
two quarks may couple to spin 0 or to spin 1, thus forming a 
mixed-antisymmetric or mixed-symmetric spin state. One main
idea of the present model is to identify the mixed-antisymmetric part
of the spin  
function with the scalar diquark channel and the mixed-symmetric one
with the axial-vector diquark channel. An axial-vector particle has
positive parity, and we consistently neglect any relative angular 
momentum between the diquark and the quark. There are indeed
theoretical arguments 
in favour of a dominance of scalar diquarks \cite{vogl,cahill},
and also of eventual axial-vector diquarks \cite{weiss,ishii}. Also, the 
amount of experimental hints of diquarks in the nucleon is not negligible,
see \cite{anselmino} for a review on that subject. On the other hand, 
there are theoretical works which clearly deny such correlations, 
see e.g. \cite{leinweber}. However, we do not want to discuss against or
in favour of diquark correlations. Obviously, this is only possible in
a full three particle calculation. Rather, we develop a simple but
formally covariant model of the nucleon which is able to reproduce
various experimental data such as electromagnetic form factors, mean
square charge radii and magnetic moments. 

As in \cite{ke} we start from the Bethe-Salpeter equation. Following
Salpeter \cite{salpeter} we assume an instantaneous interaction and
obtain a Salpeter-type equation.
As the only interaction we adopt an (instantaneous) quark exchange 
between the quark and the diquark. This interaction has been
previously used in various studies
\cite{lichtenberg,huang,reinhardt,meyer,hellstern,hanhart,ishii} and 
seems most natural for a quark-diquark model. Involving only scalar and 
axial-vector diquark channels, we deduce a pair of coupled integral
equations similar to \cite{meyer}. Within a basis of positive parity
amplitudes with spin $\half$, we obtain a bound state solution below
the threshold for this Schr\"odinger-type equation. Then,
electromagnetic transition currents and form factors are calculated
using the Mandelstam formalism \cite{mandelstam}. For details of the
calculation see \cite{ke}.  

This paper is structured as follows. In Sec. \ref{model} we 
summarize the basic equations of the model. The pair of coupled
Salpeter equations is presented. The hermiticity condition for the
interaction is shortly outlined. Sec. \ref{formfactors} presents the
idea of how the proton and 
neutron currents are calculated. The results of some current matrix
elements and of the form factors are given in Sec. \ref{results}. We
compare with experimental data. Finally, a summary is given in
Sec. \ref{summary}.

\section{The Model} 
\label{model}

\subsection{The diquark-quark Salpeter equation}
\label{equation}

A relativistic bound state of a scalar or (axial-)vector particle 
and a quark with total four momentum $P$ is described by
a Bethe-Salpeter amplitude 
\bqn
\chi_P(x_1,x_2)_{(\mu) \alpha} = 
\bramket{0}{T \; \phi_{(\mu)}(x_1) \psi_\alpha(x_2)}{P} \; .
\eqn
In momentum space, $\chi_P$ fulfills the following Bethe-Salpeter
equation \cite{BS}
\bqn \label{BSequation}
\chi_P(p)_{(\mu)} = \Delta_1^F(p_1)_{(\mu)}^{\;\; (\nu)} S_2^F(p_2) 
            \intpp{p'} \left( -i K(P,p,p') \chi_P(p') \right)_{(\nu)}
\eqn
with $\Delta_1^F(p_1)_{(\mu)}^{\;\;(\nu)}$ and $S_2^F(p_2)$ the Feynman 
propagators for a scalar (vector) particle and a quark, respectively, 
and $-i K$ the irreducible interaction kernel.  
The fundamental equation of our model has been derived in 
\cite{ke}. It is a Schr\"odinger-type equation of the 
following form (all indices suppressed):
\bqn \label{salpeter}
({\cal H} \Psi)(\vec p\,) & = & M \; \Psi(\vec p\,) \nn
& = & \frac{\omega_1+\omega_2}{\omega_2} \; H_2(\vec p\,) \; \Psi(\vec p\,)
      +\frac{1}{2 \omega_1} \intp{p'} \;  
       W(\vec p, \vec p\,') \; \Psi(\vec p\,') \nn
& = & ({\cal T} + {\cal W}) \; \Psi (\vec p\,) \; . 
\eqn
$\Psi(\vec p\,)$ is the Bethe-Salpeter amplitude in the bound state's rest frame 
integrated over $p^0$:
\bqn
\Psi(\vec p\,) = \gamma^0 
\left( \int \frac{d p^0}{2\pi} \chi_P(p^0,\vec p\,) \right)
_{P=(M,\vec 0)}  \; .
\eqn
$W(\vec p, \vec p\,') = K(P,p,p') \gamma^0$ is the interaction kernel
in the rest frame
which is supposed to be independent of $p^0$ and $p'^{\;0}$.
$H_2(\vec p\,)$ is the Dirac Hamiltonian of the quark, and 
$\omega_i^2 = \vec p\,^2+m_i^2$ the squared on-shell energies of the diquark 
and quark, respectively. The masses $m_i$ are constituent masses of
the order of a few hundred MeV, see Sec. \ref{results}. $M$ is the
mass of the bound state. For the normalization we obtained in \cite{ke}:
\bqn \label{norm}
\braket{\Psi}{\Psi} & = & 2 M   
\eqn
with the scalar product defined as
\bqn \label{scalarproduct}
\braket{\Psi'}{\Psi} & = & 
\intp{p} (2 \omega_1) \; \Psi'^\dagger(\vec p\,) \; \Psi(\vec p\,) \; . 
\eqn
To obtain real eigenvalues $M$ of the pseudo-Hamiltonian 
${\cal H}$ in Eq. (\ref{salpeter}) the interaction kernel has 
to be chosen such that ${\cal H}$ is hermitian with respect to the 
scalar product (\ref{scalarproduct}).

\subsection{The interaction kernel}      
\begin{figure}
\begin{center}
\leavevmode
\epsfxsize=0.80\textwidth
\epsffile{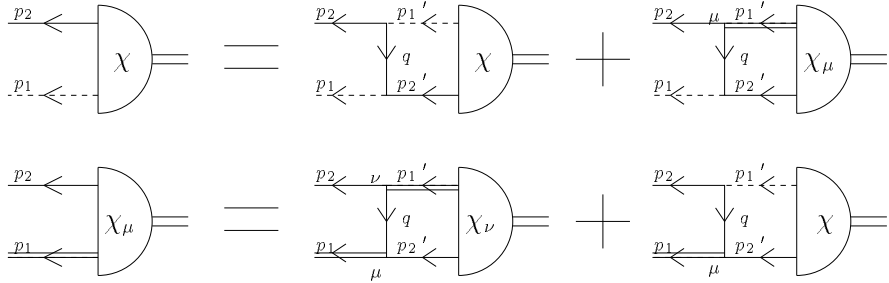}
\end{center}
\caption{The (Bethe-)Salpeter equations for a bound state of a quark and
  a scalar or axial-vector diquark with a quark exchange interaction,
  see Eq. (\protect\ref{quex}). The dashed line represents a scalar
  diquark and the double-lined an axial-vector diquark.}
\label{BSfig}
\end{figure}

Scalar and axial-vector diquarks (called v-diquarks in the following)
couple to two quarks 
according to the Lagrangians \cite{meyer}
\bqn \label{lagrangian}
{\cal L}_{int}^{scalar} & = & -i \; g_s \; \ov\psi_c \; \gamma^5 \; 
\tau_2 \; \psi \; \phi^*  \; , \nn
{\cal L}_{int}^{axial-vector} & = & - g_v \; \ov\psi_c \; \gamma^\mu \; 
\tau_a \; \psi \; {\phi_\mu^a}^* \; ,
\eqn
respectively. Here, $g_s$ and $g_v$ are dimensionless coupling constants, 
and we will choose $g_s = g_v = g$, see Sec. \ref{results}. To obtain 
a stable solution of the Salpeter equation (\ref{salpeter}) (or 
(\ref{quex}), see below),  
we have to introduce a form factor of the diquarks, see \cite{ke}. 
It is assumed to be the same for the scalar and v-diquark and is 
chosen to be a Gaussian function. 
Again we use as the only interaction between the diquark and quark 
a quark exchange and perform the instantaneous approximation, i.e.
neglect the $p^0$ dependence. So the interaction kernel is of the form
\bqn
W(\vec p, \vec p\,') & = & 
-g^2 \frac{1}{\omega_q^2} (-\vec\gamma(\vec p+\vec p\,') + m_q)
\gamma^0 \; ,
\eqn
with $\omega_q = \sqrt{(\vec p + \vec p\,')^2+m_q^2}$ the energy of
the exchanged quark. We redefine the above $g$ by absorbing the colour
factor of 2 (For a given colour of the diquark there are two possible
colours of the exchanged quark). We then obtain from Eq. (\ref{salpeter}) the
following coupled integral equations for the scalar and v-diquark
components of the nucleon, namely $\Psi^{[0]}$ and $\Psi_\mu^{[1]}$
(see Fig. \ref{BSfig}) (without flavour dependence):
\bqn 
\label{quex}
M \Psi^{[0]}(\vec p\,) & = &
\frac{\omega_1+\omega_2}{\omega_2} H_2(\vec p\,) \Psi^{[0]}(\vec p\,) \nn
& & +\frac{1}{2 \omega_1} \intp{p'} (-g_s^2) \frac{1}{\omega_q^2}
(- \vec \gamma (\vec p+\vec p\,')+m_q)
\gamma^0 \Psi^{[0]}(\vec p\,') \nn
& & +\frac{1}{2 \omega_1} \intp{p'} (-g_s g_v) \frac{1}{\omega_q^2}  
\gamma^\mu \gamma^5 (- \vec \gamma (\vec p+\vec p\,')+m_q) 
\gamma^0 \Psi_\mu^{[1]}(\vec p\,') \; , \nn 
M \Psi_\mu^{[1]}(\vec p\,) & = & 
\frac{\omega_1+\omega_2}{\omega_2} H_2(\vec p\,) \Psi_\mu^{[1]}(\vec p\,) \nn
& & +\frac{1}{2 \omega_1} \intp{p'} (+g_v^2) \frac{1}{\omega_q^2}
\gamma^\nu \gamma^5 (- \vec \gamma (\vec p+\vec p\,')+m_q)
\gamma^5 \gamma_\mu^\dagger \gamma^0 \Psi_\nu^{[1]}(\vec p\,') \nn
& & +\frac{1}{2 \omega_1} \intp{p'} (+g_s g_v) \frac{1}{\omega_q^2}  
(-\vec \gamma (\vec p+\vec p\,')+m_q) 
\gamma^5 \gamma_\mu^\dagger \gamma^0 \Psi^{[0]}(\vec p\,') \; . 
\eqn
The above interaction ${\cal W}$ (compare with Eq. (\ref{salpeter})) is indeed
hermitian with respect to the scalar product (\ref{scalarproduct}),
since the following conditions are fulfilled:
\bqn
\begin{array}{rcl}
W^{[0][0]}(\vec p, \vec p\,')^\dagger & = & 
W^{[0][0]}(\vec p\,', \vec p\,) 
\vspace{0.1cm} \\
{W^{[1][1]}}^\mu_{\;\;\nu} (\vec p, \vec p\,')^\dagger & = & 
{W^{[1][1]}}^\nu_{\;\;\mu} (\vec p\,', \vec p\,) 
\vspace{0.1cm} \\
{W^{[1][0]}}_\mu (\vec p, \vec p\,')^\dagger & = & 
{W^{[0][1]}}^\mu (\vec p\,', \vec p\,) \; ,
\end{array}
\eqn
where e.g. 
\bqn
{W^{[1][0]}}_\mu (\vec p, \vec p\,') & = & 
(+g_s g_v) \frac{1}{\omega_q^2} (-\vec \gamma (\vec p+\vec p\,')+m_q) 
\gamma^5 \gamma_\mu^\dagger \gamma^0  \; .
\eqn
The sign of $W^{[0][0]}$ is chosen in \cite{ke} to obtain a bound
state of only a scalar diquark and a quark. With respect to a further
extension of our model (see Sec. \ref{summary}) we choose the sign of
$W^{[1][1]}$ to be positive. It can be shown that this sign gives rise
to a bound $\Delta(3/2)$ composed of a v-diquark and a quark.

\subsection{Solving the diquark-quark equation}

The procedure solving Eq. (\ref{quex}) is analogous to \cite{ke}
and involves the Ritz variational principle. As a basis of the nucleon 
with spin ($\half$, s) in spin and Lorentz space we choose
\bqn 
\label{basis}
{e_s^i}^{[0]} & = & {}^t(e_s^i,0,0) \; , \; 
{e_s^i}_{\;0}^{[1]} =  {}^t(0,e_s^i,0) \; , \; 
{e_s^i}_{\;V}^{[1]} = {}^t(0,0,e_s^i) \; , \nn
\mbox{with } \; \; e_s^1(\hat p) & = & \zweivek{\chi_s}{0} \; \; , \; \;
e_s^2(\hat p) =  \zweivek{0}{\frac{\vec \sigma \hat p}{W+M_{1 2}}
  \chi_s} \; ,
\eqn
where $M_{1 2} = \frac{m_1 m_2}{m_1+m_2}$ and 
$W^2 = \vec p \,^2+M_{12}^2$. 
Note that we combined the three vector components of the 
v-diquark to one reduced function only carrying total spin 1. 
The corresponding basis vector is denoted ${e_s^i}_{\;V}^{[1]}$. The
contribution from the zero-component of the v-diquark 
is expanded in terms of ${e_s^i}_{\;0}^{[1]}$ and that from the
scalar diquark channel in terms of ${e_s^i}^{[0]}$. 
However, the basis states of Eq. (\ref{basis}) are not those involved
when calculating the matrix elements 
of the potential in Eq. (\ref{quex}). There, the $\gamma$ matrices
are coupled with the {\em quark} spin. Thus, for the quark coupling 
to the zero-component
of the v-diquark (via $\sim \gamma^5 \gamma^0$), also the following
basis states with {\em negative} parity  
are needed (see App. \ref{appparity}, \ref{applorentz}):
\bqn
e_s^3(\hat p) & = & \zweivek{0}{\chi_s} \;\; , \; \;
e_s^4(\hat p) = \zweivek{\frac{\vec \sigma \hat p}{W+M_{1 2}} \chi_s}{0} 
\eqn 
Then,
\bqn
\begin{array}{rcl}
\gamma^5 \, \gamma^0 \, e_s^3(\hat p) & \sim & e_s^1(\hat p) 
\vspace{0.1cm} \\
\gamma^5 \, \gamma^0 \, e_s^4(\hat p) & \sim & e_s^2(\hat p) 
\vspace{0.1cm} \\
\gamma^5 \, \left[ \gamma^{[1]} \otimes e^1(\hat p) \right]_s^\half 
& \sim & e_s^1(\hat p) 
\vspace{0.1cm} \\
\gamma^5 \, \left[ \gamma^{[1]} \otimes e^2(\hat p) \right]_s^\half 
& \sim & e_s^2(\hat p)
\end{array}
\eqn
are basis states of positive parity, see also App. \ref{appparity}.

\section{Current matrix elements and form factors}
\label{formfactors}
\begin{figure}
\begin{center}
\leavevmode
\epsfxsize=0.85\textwidth
\epsffile{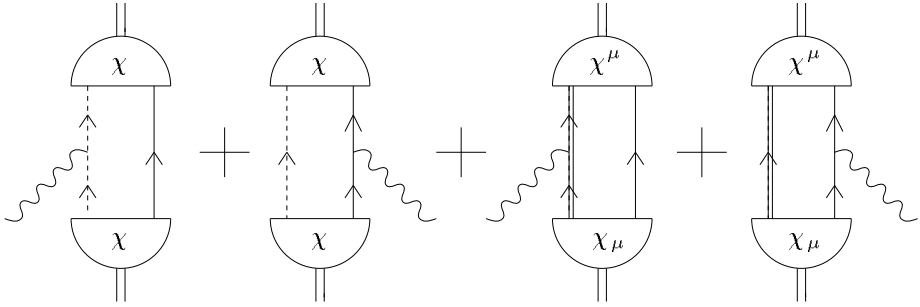}
\end{center}
\caption{The electromagnetic current is the sum of the diquark currents
  and the quark currents.}
\label{currentfig}
\end{figure}

\begin{figure}
\begin{center}
\leavevmode
\epsfxsize=0.30\textwidth
\epsffile{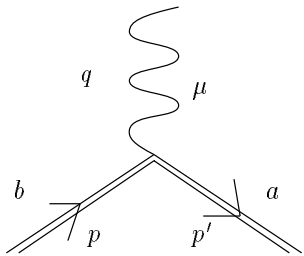}
\end{center}
\caption{The coupling of the photon to a massive vector particle
  (Eq. (\protect\ref{photonvector})).} 
\label{couplfig}
\end{figure}

The current matrix elements are calculated in the Mandelstam 
formalism \cite{mandelstam}. The corresponding diagrams are 
shown in Fig. \ref{currentfig}. For details of the calculation see
\cite{ke}. The Bethe-Salpeter amplitudes of course depend on 
the total and relative momenta $P$ and $p$. A correct boost
prescription (App. \ref{applorentz}) for the outgoing $\ov \chi$ is crucial
for a relativistic treatment. Since the implementation of
this formalism in the scalar diquark-quark model is elaborated in 
\cite{ke} we only focus our attention on aspects 
concerning the v-diquark. \\
The propagator of the v-diquark is chosen to be
\bqn
\label{vprop}
\Delta_1^F(p)_{\mu \nu} = - i \frac{g_{\mu \nu}}{p^2-m^2+ i \eps} \; ,
\eqn 
i.e. we neglect the term 
$\frac{p_\mu p_\nu/m^2}{p^2-m^2 + i \eps}$. 
There are several technical reasons for this: 
Firstly, it is not possible to state a coordinate independent
normalization condition with the full propagator. Secondly, in 
the Salpeter approach, the contour  
integral over the full propagators in the complex plane does not vanish
at infinity for the components with $\mu, \nu = 0$. Finally, the full
propagator has no inverse, which, however, is mandatory in the
Salpeter approach, e.g. to derive the normalization condition. \\
The coupling of a photon with index $\mu$ to a massive vector particle
is given by  
\cite{leeyang,kroll} (see Fig. \ref{couplfig}):
\bqn \label{photonvector}
\Gamma_{\mu; \; b a} & = & - (p + p')_\mu \; g_{b a} 
+ ((1+\kappa) p_b - (\kappa+\xi) p'_b) \; g_{\mu a} \nn
& & + ((1+\kappa) p'_a - (\kappa+\xi) p_a) \; g_{\mu b} \; ,
\eqn
with $\kappa$ the anomalous contribution to the magnetic moment
of the v-diquark ($\kappa = 1$ for a pointlike spin--1
particle). $\xi$ is a gauge parameter introduced
by Lee and Yang. According to our choice of the propagator (\ref{vprop})
we put $\xi = 1$. We will see that this is indeed necessary to obtain
a conserved current and the correct normalization of the Salpeter
amplitudes. \\
Rewriting the coupling (\ref{photonvector}) in spherical components with
indices $a$ and $b$ and taking into account
the corresponding Clebsch-Gordan coefficients, we obtain the
coupling matrices in the space of  
$(e^{[0]},e_{\;0}^{[1]},e_V^{[1]})$, i.e. including both scalar and v-diquark
couplings:
\begin{eqnar} \label{coupling}
\label{coupling1}
\Gamma^0 & = & \dreimat{(p_1+p'_1)^0} {0} {0}
                     {0} {-(p_1+p'_1)^0} 
                     {\frac{1}{\sqrt{3}}(1+\kappa) \, q}
                     {0} {-\frac{1}{\sqrt{3}}(1+\kappa) \, q} 
                     {(p_1+p'_1)^0} \\
\label{coupling2}
\Gamma^3 & = & \dreimat{-(p_1+p'_1)^3}{0}{0}
                     {0} {(p_1+p'_1)^3} 
                     {\frac{1}{\sqrt{3}}(1+\kappa) \, q^0}
                     {0} {-\frac{1}{\sqrt{3}}(1+\kappa) \, q^0}
                     {-(p_1+p'_1)^3} \\
\label{coupling3}
\Gamma^+ & = & \dreimat{0} {0} {0}
                     {0} {0} {\frac{1}{\sqrt{3}}(1+\kappa) \, q^0}
                     {0} {-\frac{1}{\sqrt{3}}(1+\kappa) \, q^0}
                     {-\frac{2}{3}(1+\kappa) \, q} \; ,      
\end{eqnar}  
where $\Gamma^+ = \half (\Gamma^1+i\Gamma^2)$. $(q^0, \vec q \,)$ is
the photon momentum and $q = |\vec q \,| = - q^3$. 
Transitions between scalar and v-diquarks are neglected 
since they would change the flavour symmetry in our model
\cite{hellstern,kroll}. 
From Eq. (\ref{coupling1}) we see that, in accordance
with the choice of the propagator, the zero component of the v-diquark 
contributes negatively to the norm. As in \cite{ke} we see that the
current is conserved exactly also in the v-diquark channel. \\
In the non-relativistic quark-model the spin-flavour wave function
of the nucleon for zero angular momentum is totally-symmetric, 
see App. \ref{wavefunct}.
We identify 
\bqn \label{spinwavefunct}
\begin{array}{ccc}
{\cal R}(p) \left[ \left[ \chi^\half \otimes \chi^\half \right]^0
  \otimes \chi^\half \right]_s^\half 
& \ra & \Psi^{[0]}(\vec p\,) 
\vspace{0.1cm} \\
{\cal R}(p) \left[ \left[ \chi^\half \otimes \chi^\half \right]^1
  \otimes \chi^\half \right]_s^\half 
& \ra & \Psi^{[1]}(\vec p\,) \; . 
\end{array}
\eqn
One then obtains for the current matrix elements of the proton and
neutron, respectively 
(see App. \ref{wavefunct}):
\bqn \label{pncurrent}
\begin{array}{ccl}
\bramket{\Psi_p}{j_\mu}{\Psi_p} & = & 
\frac{1}{3}   \bramket{\Psi^{[0]}}{j_\mu^{diquark}}{\Psi^{[0]}} 
+ \frac{2}{3} \bramket{\Psi^{[0]}}{j_\mu^{quark}}{\Psi^{[0]}} 
\vspace{0.1cm} \\
& & + \bramket{\Psi^{[1]}}{j_\mu^{diquark}}{\Psi^{[1]}} 
\vspace{0.1cm} \\
\bramket{\Psi_n}{j_\mu}{\Psi_n} & = & 
\frac{1}{3}   \bramket{\Psi^{[0]}}{j_\mu^{diquark}}{\Psi^{[0]}}
- \frac{1}{3} \bramket{\Psi^{[0]}}{j_\mu^{quark}}{\Psi^{[0]}} 
\vspace{0.1cm} \\
& & - \frac{1}{3} \bramket{\Psi^{[1]}}{j_\mu^{diquark}}{\Psi^{[1]}} 
    + \frac{1}{3} \bramket{\Psi^{[1]}}{j_\mu^{quark}}{\Psi^{[1]}} \; . 
\end{array}
\eqn
Note that in the case of the proton, the photon coupling to the quark
in the v-diquark channel drops out, 
whereas in the case of the neutron the current is sensitive to the 
difference between the quark and the diquark coupling.

\section{Results and discussion}
\label{results}
\begin{figure}
\begin{center} 
\leavevmode
\epsfxsize=0.65\textwidth
\epsffile{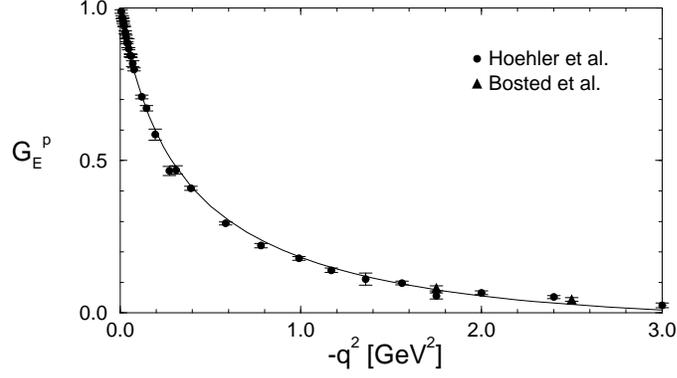}
\end{center}
\caption{The electric form factor of the proton $G_E^p(q^2)$;  
  experimental data are taken from \protect\cite{hoehler,bosted}.}
\label{geproton}
\end{figure}

\begin{figure}
\begin{center} 
\leavevmode
\epsfxsize=0.65\textwidth
\epsffile{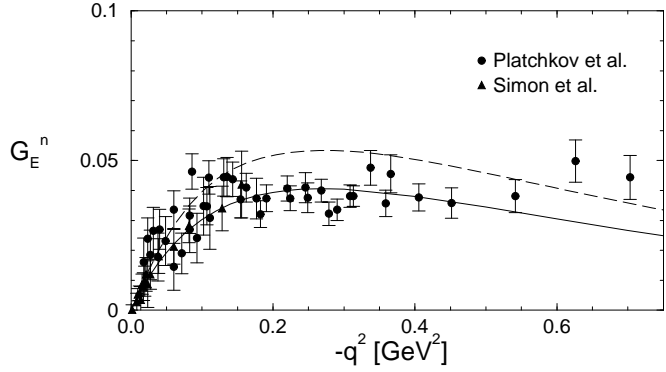}
\end{center}
\caption{The electric form factor of the neutron $G_E^n(q^2)$; 
  experimental data are taken from \protect\cite{simon,platchkov}. The
  full line corresponds to the parameter set of
  Tab. \protect\ref{param}, the dashed line corresponds to 
  $m_v = 670 \; $ MeV.}
\label{geneutron}
\end{figure}

\begin{figure}
\begin{center} 
\leavevmode
\epsfxsize=0.65\textwidth
\epsffile{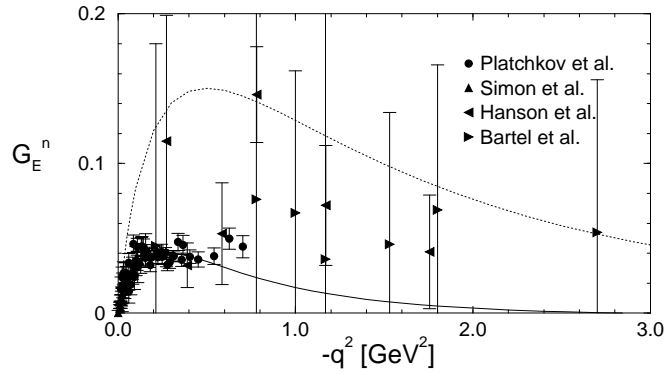}
\end{center}
\caption{The electric form factor of the neutron $G_E^n(q^2)$; 
  experimental data are taken from
  \protect\cite{simon,platchkov,hanson,bartel}. The
  full line corresponds to the parameter set of
  Tab. \protect\ref{param}, the dotted line is our calculation with
  scalar diquarks only (from \protect\cite{ke}).}
\label{geneutronall}
\end{figure}

\begin{figure}
\begin{center} 
\leavevmode
\epsfxsize=0.65\textwidth
\epsffile{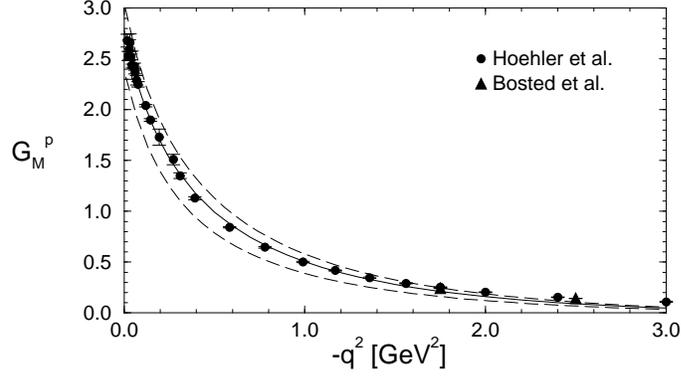}
\end{center}
\caption{The magnetic form factor of the proton $G_M^p(q^2)$;  
  experimental data are taken from \protect\cite{hoehler,bosted}.
  The full line corresponds to the parameter set of 
  Tab. \protect\ref{param}, the upper and lower dashed curve to
  $\kappa = 2.0$ and $\kappa = 1.0$, respectively.}
\label{gmproton}
\end{figure}

\begin{figure}
\begin{center} 
\leavevmode
\epsfxsize=0.65\textwidth
\epsffile{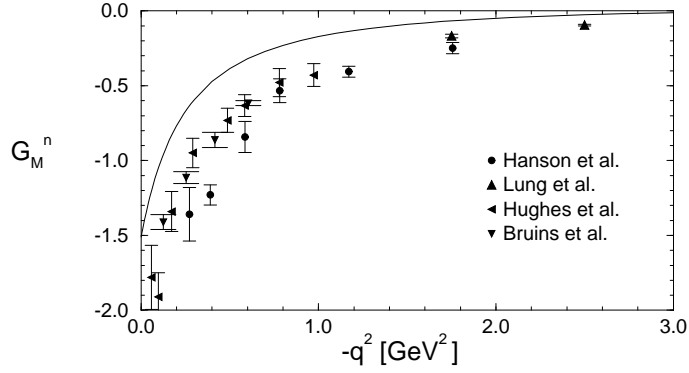}
\end{center}
\caption{The magnetic form factor $G_M^n(q^2)$ of the neutron; 
  experimental data are taken from
  \protect\cite{hanson,lung,hughes,bruins}.} 
\label{gmneutron}
\end{figure}

\begin{figure}
\begin{center} 
\leavevmode
\epsfxsize=0.65\textwidth
\epsffile{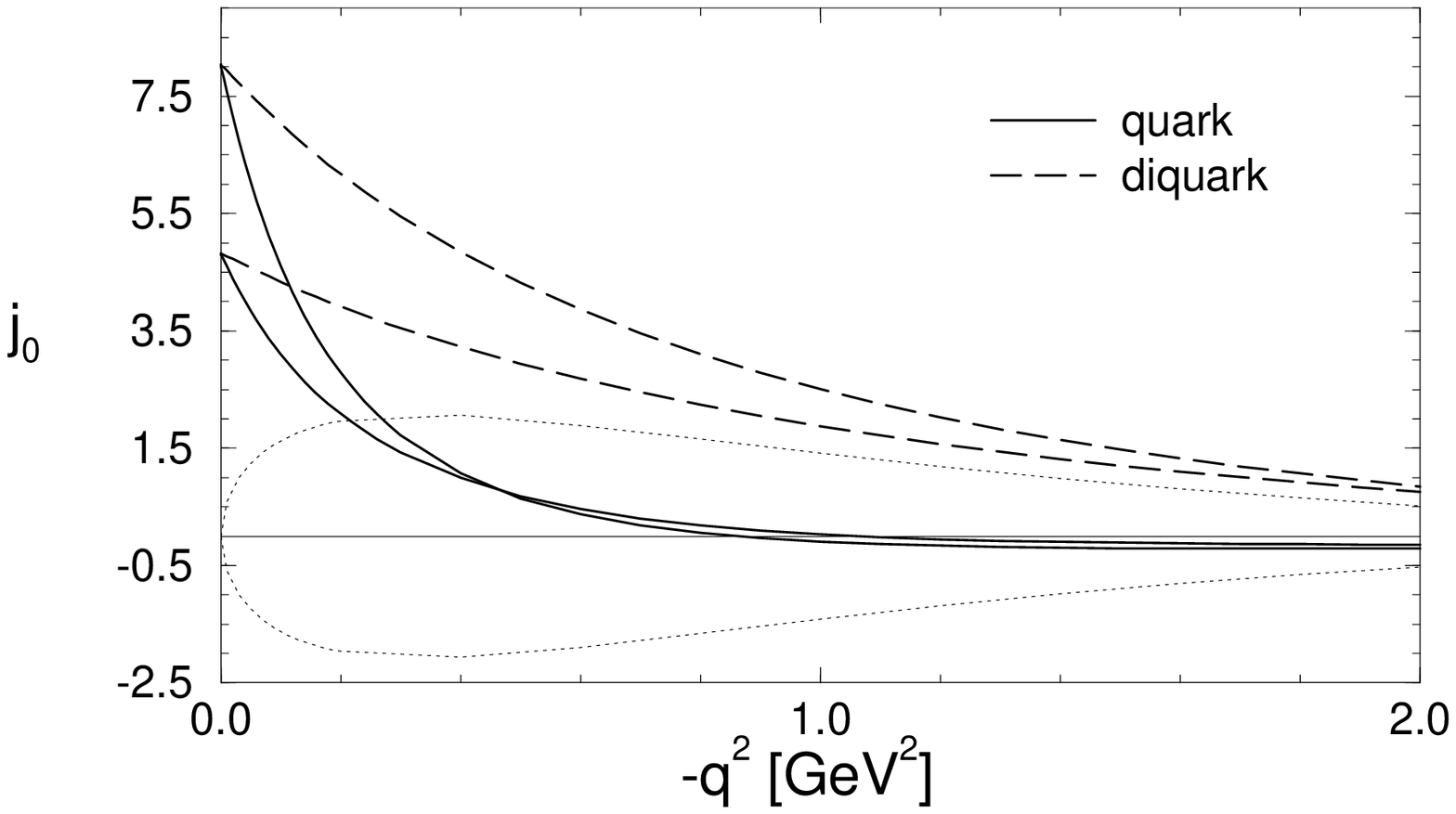}
\end{center}
\caption{The charge density $j_0(q^2)$ of the four currents of
  Fig. \protect\ref{currentfig} in dimensionless units. The
  full lines describe the coupling to the quark, the dashed lines that
  to the diquark. The upper pair of full and dashed lines corresponds
  to the currents in the scalar diquark channel, the lower one to
  those in the v-diquark channel (with zero- and vector components added). 
  The dotted lines correspond to the non-diagonal couplings
  of Eq. (\protect\ref{coupling1}).}   
\label{jmunull}
\end{figure}

\begin{figure}
\begin{center} 
\leavevmode
\epsfxsize=0.65\textwidth
\epsffile{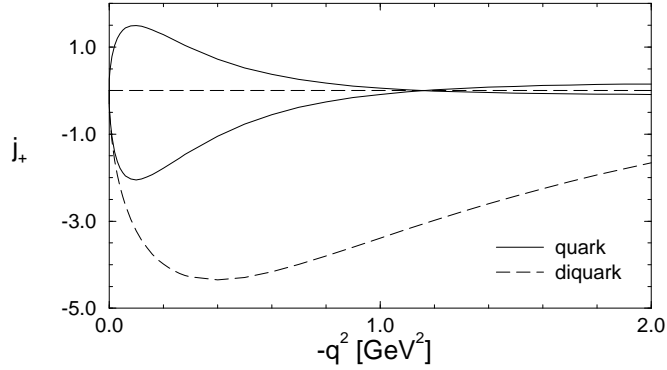}
\end{center}
\caption{The spin-flip currents $j_+ = \half (j_1 + i \; j_2)$ (see
  Fig. \protect\ref{currentfig}) in dimensionless units. The full
  lines describe the coupling 
  to the quark, the dashed lines that to the diquark. The full line
  starting with negative values is the quark current in the scalar diquark
  channel. The $j_+$ current of the scalar diquark is trivially zero.} 
\label{jmuplus}
\end{figure}

The values of the parameters used in our model are given in 
Tab.\ref{param}. The quark and diquark masses are the same as in our 
previous work \cite{ke}. The mass of the v-diquark is chosen to be
equal to the scalar diquark mass. Also, the quark-diquark coupling
constants are chosen to be equal: $g = g_s = g_v$. 
Their value is fixed by the minimum of the $M(\alpha)$ curve in the 
Ritz variational principle, analogous to \cite{ke}. For six radial
basis functions we have a minimum at $M(\alpha) = 939 \; $ MeV for the
oscillator parameter $\alpha = 1.35 \;$ fm. With equal masses and
couplings we are closest to the $SU(2)_{spin} \otimes SU(2)_{isospin}$
limit which is
only broken by the different types of coupling ($\sim 1$ for scalar and 
$\sim \gamma^\mu$ for the v-diquark) (Eqs. (\ref{lagrangian}) and
(\ref{quex})).   
Indeed, we find in this most symmetric case a best description of all
electromagnetic form factors. 
The parameter influencing the shape of the form factors mostly is the 
diquark form factor parameter $\lambda$, see \cite{ke}. It is put equal
for both diquark types and is chosen to give a best description of the
electric form factor of the proton. We obtain $\lambda = 0.24 \; $ fm,
in agreement with the commonly used diquark size \cite{anselmino}. The
proton electric form factor is shown in Fig. \ref{geproton} for
momentum transfers up to $- q^2 = 3 \; \mbox{ GeV}^2$. We find a very
good agreement with the experimentally found dipole shape. However, 
for about $- q^2 > 3.5 \; \mbox{ GeV}^2$, $G_E^p$ becomes
negative, though with small absolute value and converging to
zero. This change of sign is the case for all four form factors
and is model inherent. From the
slope at $q^2 = 0$ we get for the rms-radius of the proton 
\bqn
\sqrt{\langle r^2 \rangle_p} = 0.84 \; \mbox{ fm} \; ,
\eqn
compared to the experimental 
$\sqrt{\langle r^2 \rangle_p} = (0.862 \pm 0.012 )\; $ fm
\cite{simon}.
In Fig. \ref{geneutron} the neutron electric form factor is 
shown for momentum transfers up to $- q^2 = 0.75 \; \mbox{ GeV}^2$
where $G_E^n$ is well established experimentally. We obtain a
remarkable description of the experimental data.  
For the mean square charge radius of the neutron we find 
\bqn
\langle r^2 \rangle_n & = & - 0.118 \; \mbox{fm}^2 \; ,
\eqn
which fits the experimental 
$\langle r^2 \rangle_n = -0.119 \pm 0.002 \; \mbox{fm}^2$
\cite{bostedprc}. Since the neutron current is the 
difference of currents (see Eq. (\ref{pncurrent})), it is very sensitive 
to the parameters. The dashed curve shows $G_E^n$
for a v-diquark mass $m_v = 670 \;$ MeV. Indeed, we describe the data
well in the mass-symmetric case. In Fig. \ref{geneutronall} we compare
our results (full curve) with our recent calculation only involving
scalar diquarks \cite{ke} (dotted curve). The improvement is obvious. \\
Fig. \ref{gmproton} shows the magnetic form factor of the proton.
As can be seen from Eq. (\ref{coupling3}), the spin flip current of
the v-diquark is proportional 
to $(1+\kappa)$, $\kappa$ being the anomalous magnetic moment of the
v-diquark. This is chosen
to be $\kappa = 1.6$ to fit the experimental data for $G_M^p$. Again,
our calculation reproduces the dipole shape very nicely. 
The dependence of $G_M^p$ on $\kappa$ is shown by the dashed
curves. The upper dashed curve corresponds to $\kappa = 2.0$, the
lower one to $\kappa = 1.0$. The larger the diquark's anomalous
magnetic moment the larger is the absolute value of the magnetic form
factor.  
The neutron magnetic form factor is shown in Fig. \ref{gmneutron}. The
calculation matches the experimental data qualitatively. However, the
absolute value of the calculated curve is too small.
Extrapolating to $q^2 = 0$, we get from $G_M^N(0)$ for the magnetic
moments of the proton and neutron: 
\bqn
\begin{array}{rccc}
\mu_p & = & 2.78 \; \mu_N 
 & \mbox{(exp.: } \; 2.793 \; \mu_N \mbox{)} 
\vspace{0.1cm} \\
\mu_n & = & - 1.51 \; \mu_N 
& \mbox{(exp.: } \; -1.913 \; \mu_N \mbox{)} 
\vspace{0.1cm} \\
\ra \;\; \frac{\mu_n}{\mu_p} & = & - 0.543 
& \mbox{(exp.: } \; -0.685 \mbox{)} \; .  
\end{array}
\eqn
The $SU(6)$ quark model prediction is $-\frac{2}{3}$. \\ 
In Fig. \ref{jmunull} we show the zero-components of the various currents 
(see Fig. \ref{currentfig}). As shown analytically in \cite{ke}, 
the corresponding quark and diquark currents have to be equal
at $q^2 = 0$, thus equally contributing to the normalization of the Salpeter
amplitude (Eq. (\ref{norm})). 
It is interesting to compare these contributions of the three channels
(scalar : zero-component of v-diquark : vector-component of v-diquark):
\bqn
 0.63 : -0.23 : 0.60 \; ,
\eqn
which reproduces the result of \cite{meyer} in the mass-symmetric case
(where both components of the v-diquark-channel are added): 
$0.64 : 0.36$. \\
The four spin flip currents are shown in Fig. \ref{jmuplus}. It is mainly
the current $\bramket{\Psi^{[1]}}{j_+^{diquark}}{\Psi^{[1]}}$ (lowest
dashed curve) which improves our results for the magnetic form factors
compared to our recent work \cite{ke}. It is interesting to see that
at $- q^2 > 1 \; \mbox{ GeV}^2$ the spin flip current, i.e. the magnetic
form factor, is essentially due to the spin flip of the v-diquark. 
The transitions
between the zero-component and the vector-component of the v-diquark, 
corresponding to the non-diagonal elements in the coupling matrices of
the Eqs. (\ref{coupling1})-(\ref{coupling3}) are seen to add up to zero:
\bqn
\bramket{\Psi_{\;0}^{[1]}}{j_\mu^{diquark}}{\Psi_V^{[1]}} & = & 
- \bramket{\Psi_V^{[1]}}{j_\mu^{diquark}}{\Psi_{\;0}^{[1]}} \; ,
\eqn  
see the dotted curves in Fig. \ref{jmunull} for $\mu = 0$.

\section{Summary and outlook}\label{summary}
We developed a formally covariant constituent quark-diquark model of
the nucleon. Two kinds of diquarks, scalar and axial-vector ones, are
taken into account. The only interaction considered is a quark
exchange. Starting from the Bethe-Salpeter equation for a bound state
of a quark and a massive boson, we otain in the instantaneous
approximation for the interaction kernel a pair of coupled
Salpeter-equations. As a further approximation we choose a
scalar-vector symmetric form of the spin-1 propagator. The equations
are solved involving the Ritz variational principle with a finite
number of radial basis functions. To obtain a stable solution, a form
factor of the diquark has to be introduced. The current matrix elements are
calculated in the Mandelstam formalism. With five parameters in the
scalar-vector symmetric case (constituent quark and diquark masses,
diquark size parameter $\lambda$, diquark-quark coupling $g$ and the
anomalous magnetic moment of the axial-vector diquark) we find an
excellent agreement with the experimental electromagnetic form factors
for momentum transfers up to $- 3 \; \mbox{ (GeV/c)}^2$. Only the
magnetic form factor of the neutron is a little off the experimental
curve. The calculated charge radii agree with the experimental ones,
and we find the correct anomalous magnetic moment of the proton. 

It is indeed surprising to find such a good correspondance with the
experimental data in such a simple model. To what extent this
is due to the chosen quark-exchange interaction, which is essentially
$\sim \frac{1}{q}$ ($\sim \frac{1}{r^2}$ in coordinate space), remains to be
examined. One important ingredient is the covariant treatment and the
correct boosting of the outgoing amplitudes. In what sense our results
may provide a revealing of a probable diquark-structure in the nucleon
is unclear. In any case, by introducing an effective potential and
diquark parameters, the complicated three-body problem can be reduced
to an effective two-body one. 

A further important test of our model is the calculation of N-$\Delta$
transition form factors. In our model only the axial-vector diquark
channel will contribute to this process. So the involved parameters
should be further constrained. \\

{\bf Acknowledgements:} I am grateful to H.R. Petry, B.C. Metsch and
C.R. M\"unz for many helpful discussions and T. Schirmer for reading
the manuscript. This work was supported by the Graduiertenkolleg 
'Die Erforschung subnuklearer Strukturen der Materie'. 

\newpage

\begin{table}
\begin{tabular}{cccccc}
$m_q$ & $m_s$ & $m_v$ & $g$ & $\lambda$ & $\kappa$ \\  
\hline
\quad 350 MeV/c$^2$ \quad & 
\quad 650 MeV/c$^2$ \quad & 
\quad 650 MeV/c$^2$ \quad & 
\quad 8.10 \quad & 
\quad 0.24 fm \quad &
\quad 1.6 \quad \\
\end{tabular}
\caption{The parameters of the model}
\label{param}
\end{table}

\newpage

\begin{appendix}

\section{Parity transformation} 
\label{appparity}

An axial-vector field transforms under parity-operation like
\cite{itzy}
\bqn
{\cal P} A^\mu (x) {\cal P} & = & - A_\mu (\tilde x)  \nn
\tilde x^\mu & = & x_\mu \; . 
\eqn
Then, we find for the transformation of the Bethe-Salpeter amplitude:
\bqn
\chi_P(p)_\mu & = & - \pi_{\cal P} \; \gamma^0 \; 
\chi_{\tilde P}(\tilde p)^\mu \; .
\eqn

\section{Lorentz transformation} 
\label{applorentz}

For the Lorentz transformation of the v-diquark-quark Bethe-Salpeter
amplitude we have \cite{itzy}
\bqn
\chi_P (p)_\mu & = & {\Lambda^{-1}}_\mu^{\;\;\nu} \; S_\Lambda^{-1} 
\; \chi_{\Lambda P}(\Lambda p)_\nu \; .
\eqn
However, as a spinor field, a nucleon field transforms like
\bqn
\Psi_P(p) & = & S_{\Lambda}^{-1} \; \Psi_{\Lambda P}(\Lambda p) \; .
\eqn
This correct transformation is achieved by the covariant coupling
\bqn
\Psi_P(p) & := & \gamma^\mu \; \chi_P(p)_\mu \; .
\eqn

\section{Nucleon wave function} 
\label{wavefunct}

Following Eq. (\ref{spinwavefunct}) we write for the proton and neutron 
wavefunctions (without the totally anti-symmetric colour wavefunction)
\bqn 
\Phi_p(\vec p\,)_s & = & {\cal N} \left( \Psi_s^{[0]}(\vec p\,) \; 
\chi_{M_A}^{uud}
+ \Psi_s^{[1]}(\vec p\,) \; \chi_{M_S}^{uud} \right) \nn
\Phi_n(\vec p\,)_s & = & {\cal N} \left( \Psi_s^{[0]}(\vec p\,) \; 
\chi_{M_A}^{udd}
+ \Psi_s^{[1]}(\vec p\,) \; \chi_{M_S}^{udd} \right) \; ,
\eqn
where the flavour functions are mixed (anti-)symmetric:  
\bqn
\chi_{M_A}^{uud} & = & \chi_0^{[0]} \nn
\chi_{M_S}^{uud} & = & -\frac{1}{\sqrt{3}}(\chi_0^{[1]}-\sqrt{2}
\chi_{+1}^{[1]}) \; ,
\eqn
with e.g. $\chi_0^{[1]} = \frac{1}{\sqrt{2}}(ud+du)u$. 

\end{appendix}

\end{document}